\documentclass[sigconf]{acmart}
\usepackage{quoting,xparse}

\AtBeginDocument{%
  }

\setcopyright{rightsretained}
\copyrightyear{2026}
\acmYear{2026}
\acmConference[WomENcourage'26]{WomENcourage}{September,2026}{Sophia Antipolis, France}

\begin{document}

\title{Beyond Retrieval: Scaffolding Children's Online Learning}


\author{Diletta Micol Tobia}
\orcid{0009-0003-9247-5002}
\email{diletta.micol.tobia@usi.ch}
\affiliation{%
 \institution{Universit\`a della Svizzera Italiana}
 \city{Lugano}
  \country{Switzerland}
 }

\author{Hrishita Chakrabarti}
\email{h.chakrabarti@tudelft.nl}
\orcid{0000-0001-7083-9686}
\affiliation{%
  \institution{Delft University of Technology}
  \city{Delft}
  \country{Netherlands}}

\author{Maria Soledad Pera}
\email{M.S.Pera@tudelft.nl}
\orcid{0000-0002-2008-9204}
\affiliation{%
  \institution{Delft University of Technology}
  \city{Delft}
  \country{Netherlands}}

\author{Monica Landoni}
\orcid{0000-0003-1414-6329}
\email{monica.landoni@usi.ch}
\affiliation{%
 \institution{Universit\`a della Svizzera Italiana}
 \city{Lugano}
  \country{Switzerland}
 }

\renewcommand{\shortauthors}{Tobia et al.}

\begin{abstract}
Children increasingly turn to online information access systems that are primarily designed for the mainstream population, e.g., adults, but possess a limited understanding of how these systems work, contributing to their unstructured and ineffective search practices. This lack of knowledge can hinder their curiosity and the development of critical search skills. Grounded on the existing literature of both child-oriented Information Retrieval and Human–Computer Interaction, our work positions children as active participants in the search process, framing it as a scaffolded learning experience rather than a simple retrieval task.
\end{abstract}





\maketitle

``\textit{Tell me and I forget, teach me and I may remember, involve me and I learn.}''
\begin{flushright}
(Benjamin Franklin, 1999)
\end{flushright}

\section{Introduction}

When it comes to learning, engagement is not optional, but essential. Childhood is the best period to acquire new learning skills \cite{janacsek2012best}, when cognitive, linguistic, and reflective abilities are still developing. This is why, during this period, children at school 
are taught concepts across various topics. School exercises and inquiry activities are designed to help students acquire that knowledge. For this, children used to turn to 
books and encyclopedias. In the last two decades, however, with the advent of technology, children have increasingly switched to online Information Access Systems \textbf{(IAS)} \cite{vanderschantz2019computer}, including Search Engines \textbf{(SE)} like Google. Yet, the vast majority of the popular tools are designed primarily with adults in mind \cite{landoni2024good}, not necessarily accounting for children's distinct search behaviors stemming from their in-development cognitive, linguistic, and information literacy skills.

From the user's perspective, Child Computer Interaction \textbf{(CCI)} research identified behavioral patterns that distinguish children's information-seeking from adults. Of note, children often lack the correct skills to be independent searchers \cite{hoppe2018current,homte2022search}. 
They tend to consider information that is more familiar and consistent with their previous experience, commonly grounding their trust in familiarity rather than in critical source evaluation \cite{tobia2026all}. When confronted with open-ended tasks, children rarely compare multiple sources or move beyond the top-ranked results in Search Engine Result Pages \textbf{(SERP)}, perceiving these results as the most trustworthy \cite{livingstone2008parental}. Moreover, they tend to follow the principle of ``least effort'' \cite{zipf2016human}, leading them to stop at the first SERP result to prioritize immediacy and reduce cognitive load. When evaluating information, they also struggle to assess the credibility and relevance of retrieved content \cite{hargittai2010trust} since they rely on surface-level cues, such as visual appearance, to evaluate content \cite{flanagin2010perceived}, resulting in an uncritical selection of information \cite{chakrabarti2025online}. From the system perspective, early Information Retrieval \textbf{(IR)} research attempted to address children's search struggles by exploring how to design IR algorithms and interfaces that take into consideration the young users' distinct search needs \citep[e.g.,][]{dowie2013releashed,bilal2002children,azpiazu2017online}. Proposed solutions, however, focused on designing new child-tailored systems that not only suffered from a lack of adoption amongst the target population \cite{bilal2012ranking}, who continue to prefer and use mainstream SE, but more importantly failed to address the core issue: children's limited search competency.

Despite the limitations documented in IR and CCI literature, SE continue to prioritize telling over involving.
To further compound the problem, children now also turn to Large Language Model (\textbf{LLM}) agents, voice-controlled assistants, and social media platforms for information seeking \cite{druin2010children}. The interaction modes offered by these tools seem to be especially helpful to overcome challenges introduced by traditional SE \cite{white2024advancing}. 
By providing immediate, fluent answers, they reduce the need for exploration, comparison, and iterative refinement. 
We argue for a change of perspective in how children's information-seeking is conceptualized: from a mere process of retrieving information to a process of learning through search. Thus, we ask a simple but fundamental question: \textit{How can we move from IAS that simplify access for children to those that actively help them become independent searchers over time?} Through the lens of the Search-as-Learning \textbf{(SAL)} paradigm \cite{collins2017search}, search becomes a process through which knowledge is actively constructed. Concepts such as uncertainty, exploration, and gradual sense-making are fundamental to learning during search \cite{kuhlthau1991inside}. When these dimensions are minimized, search risks becoming a passive act of information consumption rather than an opportunity for cognitive development.

\section{From Tasks to Learning Experiences}
To address the shift towards search as a learning process, we introduce \texttt{SOL} \texttt{(Scaffolding to Foster Independence when Children Search Online for Learning)}, an international research project that investigates how to aid children not only in finding information online, but in learning through search. In \texttt{SOL}, we reframe search as a learning activity in which the process is as important as the outcome, and where interaction itself becomes a site of cognitive development. At the core of \texttt{SOL} is the concept of scaffolding \cite{walqui2006scaffolding}, which describes the structured, adaptive support provided to learners to help them perform tasks that would otherwise be beyond their independent capabilities. This support is not meant to replace the learner’s effort, but to sustain it, gradually fading as competence develops \cite{walqui2006scaffolding}. In classroom settings, teachers already act as scaffolds for inquiry. In \texttt{SOL}, this principle is extended to digital environments, where scaffolding becomes a mechanism for preserving the exploratory nature of inquiry while making it accessible to children. 
Specifically, it is designed as an adaptive mechanism embedded within existing IAS that responds to the child’s evolving level of development, the complexity of the task, and the context of use. Depending on the child's needs, the support may include helping children formulate queries, encouraging comparison across multiple sources, providing age-appropriate explanations, or prompting reflection before accepting an answer. These examples of interventions will be designed to preserve children's agency while fostering their independence when searching online. 

We, therefore, investigate how 9-11-year-old children can be guided through their search activities, reexamining fundamental assumptions about IR for children in educational contexts. We adopt a mixed-methods approach, combining qualitative and quantitative analyses to evaluate system performance, interaction patterns, and learning outcomes. Based on iterative user testing and participatory design activities \cite{iversen2017child} with primary school children and teachers, to better understand children's information-seeking behaviors and beliefs \cite{tobia2026all}, the first objective of SOL is to define children's search profiles, or child-personas \cite{tobia2026using}, to set the ground for better personalized guidance during searching activities in the classroom. In parallel, we explore how user-IAS interactions can be enhanced through algorithmic (e.g., child-oriented retrieval and ranking strategies) and interface-level (such as visual cues and explanations) solutions to promote effective information access practices amongst children and support the evaluation of inquiry-based learning tasks. 
In SOL, we don't aim to build a new child-friendly IAS, but rather to provide a set of strategies and supporting solutions that can be embedded into existing IAS to scaffold children in their search activities. Ultimately, with \texttt{SOL}, we contribute to a shift in how we conceptualize online search for children. The goal is to move away from efficiency-driven paradigms toward learning-centered interaction design, where success is not defined by how quickly an answer is found, but by how meaningfully it is understood. In doing so, \texttt{SOL} aims to support the development of critical thinking, creativity, and digital autonomy, positioning children not as passive recipients of information, but as active participants within complex digital ecosystems, learning how to think with and beyond the information they receive for knowledge construction. By combining scaffolding practices with the principles of SAL, \texttt{SOL} advances design research through the development of age-appropriate search technologies that promote digital literacy and critical thinking. 

When children are simply told something, they may forget it; when they are taught, they may remember it; but when they are actively involved, they can truly learn. By involving them directly rather than designing for them as passive users, we posit that future generations can develop and maintain the independence and innate curiosity that characterize an active searcher.

\section{Concluding Remarks}


Children are at a critical stage where the information they encounter shapes not only how they understand the world but also how they see themselves. In an increasingly AI-mediated digital ecosystem, the \texttt{SOL} project repositions young users (an often underrepresented group) as empowered learners who actively construct knowledge through interaction with mainstream IAS, rather than passively consuming information. Grounded in the SAL paradigm and informed by scaffolding principles, our work advocates for the intentional design of IAS that promotes equitable, safe, and inclusive access to information. By rethinking algorithmic and interface-level interactions, we aim to foster children’s autonomy, critical search skills, and confidence in engaging with technology---contributing to a more diverse and equitable future in computing.


\begin{acks}
{Work supported by SNSF Award \# [IC00I0-227887 project n.10000973 SOL]}
\end{acks}

\bibliographystyle{ACM-Reference-Format}
\bibliography{references}

@article{vanderschantz2019computer,
author = {Vanderschantz, Nicholas and Hinze, Annika},
title = {“Computer what's your favourite colour?” children's information-seeking strategies in the classroom},
journal = {Proc. of the Association for Information Science and Technology},
volume = {56},
number = {1},
pages = {265-275},
year = {2019}
}

@article{kuhlthau1991inside,
  title={Inside the search process: Information seeking from the user's perspective},
  author={Kuhlthau, Carol C},
  journal={JASIST},
  volume={42},
  number={5},
  pages={361--371},
  year={1991},
  publisher={Wiley Online Library}
}

@inproceedings{iversen2017child,
  title={Child as protagonist: Expanding the role of children in participatory design},
  author={Iversen, Ole Sejer and Smith, Rachel Charlotte and Dindler, Christian},
  booktitle={Proceedings of the 2017 conference on interaction design and children},
  pages={27--37},
  year={2017}
}

@InProceedings{landoni2024good,
author="Landoni, Monica
and Huibers, Theo
and Murgia, Emiliana
and Pera, Maria Soledad",
title="Good for Children, Good for All?",
booktitle="ECIR",
year="2024",
publisher="Springer Nature Switzerland",
address="Cham",
pages="302--313",
isbn="978-3-031-56066-8"
}

@inproceedings{dowie2013releashed,
author = {Dowie, Doug and Azzopardi, Leif},
title = {Re-leashed! the PuppyIR framework for developing information services for children, adults and dogs},
year = {2013},
isbn = {9783642369728},
publisher = {Springer-Verlag},
address = {Berlin, Heidelberg},
booktitle = {Proc. of the 35th ECIR},
pages = {824–827},
numpages = {4}
}

@article{bilal2002children,
author = {Bilal, Dania},
title = {Children's use of the yahooligans! web search engine. III. cognitive and physical behaviors on fully self-generated search tasks},
year = {2002},
issue_date = {November 2002},
publisher = {John Wiley \& Sons, Inc.},
address = {USA},
volume = {53},
number = {13},
issn = {1532-2882},
journal = {J. Am. Soc. Inf. Sci. Technol.},
pages = {1170–1183},
numpages = {14}
}

@article{azpiazu2017online,
  title={Online searching and learning: YUM and other search tools for children and teachers},
  author={Azpiazu, Ion Madrazo and Dragovic, Nevena and Pera, Maria Soledad and Fails, Jerry Alan},
  journal={Information Retrieval Journal},
  volume={20},
  pages={524--545},
  year={2017},
  publisher={Springer}
}

@inproceedings{hoppe2018current,
  title={Current challenges for studying search as learning processes},
  author={Hoppe, Anett and Holtz, Peter and Kammerer, Yvonne and Yu, Ran and Dietze, Stefan and Ewerth, Ralph},
  booktitle={7th LILE Workshop, in conjunction with ACM Web Science},
  year={2018}
}

@article{homte2022search,
author = {Kameni Homte, Jaurès and Bernabé, Batchakui and Nkambou, Roger},
year = {2022},
month = {01},
pages = {254-272},
title = {Search Engines in Learning Contexts: A Literature Review},
volume = {17},
journal = {iJET}
}

@inproceedings{chakrabarti2025online,
  title={Online information disorder \& children},
  author={Chakrabarti, Hrishita and Tobia, Diletta Micol and Landoni, Monica and Pera, Maria Soledad},
  booktitle={The 5th ROMCIR Workshop (co-located with ECIR '25)},
  year={2025}
}

@InProceedings{tobia2026all,
author="Tobia, Diletta Micol
and Chakrabarti, Hrishita
and Pera, Maria Soledad
and Landoni, Monica",
title="All That Matters: Revisiting Children's Concept of Relevance in Primary School Context",
booktitle="Advances in Information Retrieval",
year="2026",
publisher="Springer Nature Switzerland",
address="Cham",
pages="271--288",
isbn="978-3-032-21324-2"
}

@article{livingstone2008parental,
  title={Parental mediation of children's internet use},
  author={Livingstone, Sonia and Helsper, Ellen J},
  journal={Journal of broadcasting \& electronic media},
  volume={52},
  number={4},
  pages={581--599},
  year={2008},
  publisher={Taylor \& Francis}
}

@book{zipf2016human,
  title={Human behavior and the principle of least effort: An introduction to human ecology},
  author={Zipf, George Kingsley},
  year={2016},
  publisher={Ravenio books}
}

@article{hargittai2010trust,
  title={Trust online: Young adults' evaluation of web content},
  author={Hargittai, Eszter and Fullerton, Lindsay and Menchen-Trevino, Ericka and Thomas, Kristin Yates},
  journal={International journal of communication},
  volume={4},
  pages={27},
  year={2010}
}

@inproceedings{flanagin2010perceived,
  title={The perceived credibility of online encyclopedias among children},
  author={Flanagin, Andrew and Metzger, Miriam},
  booktitle={Proc. of the International AAAI Conference on Web and Social Media},
  volume={4},
  number={1},
  pages={239--242},
  year={2010}
}

@inproceedings{druin2010children,
  title={Children's roles using keyword search interfaces at home},
  author={Druin, Allison and Foss, Elizabeth and Hutchinson, Hilary and Golub, Evan and Hatley, Leshell},
  booktitle={Proc. of the SIGCHI conference on human factors in computing systems},
  pages={413--422},
  year={2010}
}

@misc{white2024advancing,
      title={Advancing the Search Frontier with AI Agents}, 
      author={Ryen W. White},
      year={2024},
      eprint={2311.01235},
      archivePrefix={arXiv},
      primaryClass={cs.IR},
      url={https://arxiv.org/abs/2311.01235}, 
}

@article{collins2017search,
  title={Search as learning (dagstuhl seminar 17092)},
  author={Collins-Thompson, Kevyn and Hansen, Preben and Hauff, Claudia},
  year={2017},
  publisher={Schloss-Dagstuhl-Leibniz Zentrum f{\"u}r Informatik}
}

@article{walqui2006scaffolding,
  title={Scaffolding instruction for English language learners: A conceptual framework},
  author={Walqui, Aida},
  journal={International journal of bilingual education and bilingualism},
  volume={9},
  number={2},
  pages={159--180},
  year={2006},
  publisher={Taylor \& Francis}
}

@article{janacsek2012best,
  title={The best time to acquire new skills: Age-related differences in implicit sequence learning across the human lifespan},
  author={Janacsek, Karolina and Fiser, J{\'o}zsef and Nemeth, Dezso},
  journal={Developmental science},
  volume={15},
  number={4},
  pages={496--505},
  year={2012},
  publisher={Wiley Online Library}
}

@article{bilal2012ranking,
  title={Ranking, relevance judgment, and precision of information retrieval on children's queries: Evaluation of Google, Yahoo!, Bing, Yahoo! Kids, and ask Kids},
  author={Bilal, Dania},
  journal={JASIST and Technology},
  volume={63},
  number={9},
  pages={1879--1896},
  year={2012},
  publisher={Wiley Online Library}
}

@inproceedings{tobia2026using,
  title={Using Anti-Personas to Model Children: How to Represent User-System Mismatches},
  author={Tobia, Diletta Micol and Chakrabarti, Hrishita and Pera, Maria Soledad and Landoni, Monica},
  booktitle={Proceedings of the 34th ACM Conference on User Modeling, Adaptation and Personalization},
  pages={282--291},
  year={2026}
}

\end{document}